\documentclass[sigconf,natbib=true,anonymous=false]{acmart}

\usepackage{multirow}
\usepackage{enumerate}
\usepackage[inline]{enumitem}
\usepackage{colortbl}
\usepackage{xcolor}
\usepackage{balance}

\newcommand{\header}[1]{\noindent\textbf{#1.}}

\author{Shuyu Guo}
\affiliation{
  \institution{Shandong University}
  \city{Qingdao}
  \country{China}}
\email{guoshuyu225@gmail.com}

\author{Shuo Zhang}
\affiliation{
  \institution{Bloomberg}
  \city{London}
  \country{United Kingdom}}
\email{szhang611@bloomberg.net}

\author{Weiwei Sun}
\affiliation{
  \institution{Shandong University}
  \city{Qingdao}
  \country{China}}
\email{sunnweiwei@gmail.com}

\author{Pengjie Ren}
\affiliation{
  \institution{Shandong University}
  \city{Qingdao}
  \country{China}}
\email{renpengjie@sdu.edu.cn}

\author{Zhumin Chen}
\affiliation{
  \institution{Shandong University}
  \city{Qingdao}
  \country{China}}
\email{chenzhumin@sdu.edu.cn}

\author{Zhaochun Ren}
\authornote{Corresponding Author}
\affiliation{
  \institution{Shandong University}
  \city{Qingdao}
  \country{China}}
\email{zhaochun.ren@sdu.edu.cn}

\def\authors{Shuyu Guo, Shuo Zhang, Weiwei Sun, Pengjie Ren, Zhumin Chen, and Zhaochun Ren}

\renewcommand{\shortauthors}{Shuyu Guo et al.}

\newcommand{\code}[1]{{\ttfamily#1}}

\copyrightyear{2023} 
\acmYear{2023} 
\setcopyright{acmlicensed}
\acmConference[SIGIR '23]{Proceedings of the 46th International ACM SIGIR Conference on Research and Development in Information Retrieval}{July 23--27, 2023}{Taipei, Taiwan}
\acmBooktitle{Proceedings of the 46th International ACM SIGIR Conference on Research and Development in Information Retrieval (SIGIR '23), July 23--27, 2023, Taipei, Taiwan}
\acmPrice{15.00}
\acmDOI{10.1145/3539618.3591884}
\acmISBN{978-1-4503-9408-6/23/07}

\begin{document}
\begin{sloppypar}

\title{Towards Explainable Conversational Recommender Systems}

\begin{abstract}

Explanations in conventional recommender systems have demonstrated benefits in helping the user understand the rationality of the recommendations and improving the system's efficiency, transparency, and trustworthiness. 
In the conversational environment, multiple contextualized explanations need to be generated, which poses further challenges for explanations.
To better measure explainability in conversational recommender systems (CRS), we propose ten evaluation perspectives based on concepts from conventional recommender systems together with the characteristics of CRS.
We assess five existing CRS benchmark datasets using these metrics and observe the necessity of improving the explanation quality of CRS.
To achieve this, we conduct manual and automatic approaches to extend these dialogues and construct a new CRS dataset, namely Explainable Recommendation Dialogues (E-ReDial). 
It includes 756 dialogues with over 2,000 high-quality rewritten explanations.
We compare two baseline approaches to perform explanation generation based on E-ReDial.
Experimental results suggest that models trained on E-ReDial can significantly improve explainability while introducing knowledge into the models can further improve the performance.
GPT-3 in the in-context learning setting can generate more realistic and diverse movie descriptions. 
In contrast, T5 training on E-ReDial can better generate clear reasons for recommendations based on user preferences.
E-ReDial is available at \url{https://github.com/Superbooming/E-ReDial}.

\end{abstract}


\maketitle

\section{Introduction}
\label{sec:intro}


Recommender systems provide personalized suggestions to help users find items based on their preferences and have been widely used in various online applications. 
Explanations for recommender systems are expected to clarify why such items are recommended. 
Researchers have pointed out that appropriate explanations can help improve the recommender systems' transparency, persuasiveness, effectiveness, trustworthiness, and user satisfaction~\cite{Zhang2018ExplainableRA}. 
Thus, a growing body of work has been devoted to improving the recommendation explainability through various methods, e.g., incorporating user reviews~\cite{Chen2016LearningTR, Li2017NeuralRR}, counterfactual reasoning~\cite{Tan2021CounterfactualER, Tran2021CounterfactualEF}.


Conventional recommender systems primarily predict a user's preference over an item by analyzing their past behaviours, which can neither clarify what a user likes nor explain why a user likes an item. 
In contrast, conversational recommender systems (CRS) elicit user preferences dynamically and respond to users' needs through real-time multi-turn interactions~\cite{Gao2021AdvancesAC}.
Compared to single-turn explanation generation in conventional recommender systems, CRS needs to generate multiple contextualized explanations, which poses further challenges.
As of now, there is limited progress for explainable CRS~\cite{Jannach2020ASO}.
\citet{Wen2022EGCREG} firstly try to generate explanations on real-world CRS datasets. 
However, due to the lack of evaluation methods for CRS explanations and datasets with high-quality explanations, the performance of the generated explanations is limited and not effectively evaluated.
In this study, we focus on explainable conversation recommender systems and conduct further investigations.


We elucidate the meaning of the explanation for CRS, combing explainable recommender systems~\cite{Tintarev2015ExplainingRD} and the characteristics of CRS.
Formally, it refers to a response from CRS containing relevant details about the recommended items, e.g., the recommendation reasons, the items' descriptions and personal opinions when recommending items.
To evaluate the explanation for CRS, we propose ten evaluation perspectives. 
The effectiveness, efficiency, persuasiveness, satisfaction, scrutability, transparency, trust, and representativeness are inherited from explainable recommender systems~\cite{Balog2020MeasuringRE, Chen2022MeasuringI}. 
They are used to evaluate properties common to all recommendation systems, e.g., system transparency, recommendation accuracy, and user satisfaction.
In addition, the reasonability and coherence are derived from the characteristics of the CRS.
They measure the consistency between explanations and conversations regarding logic and language, respectively.

We measure five widely used CRS datasets using a human annotation task to verify the necessity of explainability in CRS. 
We design a questionnaire containing 13 questions covering 10 evaluation perspectives we proposed. 
With 50 dialogues sampled from each dataset, over 20 participants are invited to answer these questions at the dialogue exchange level.
According to the questionnaire results, the existing datasets have relatively low-quality explanations.
Specifically, lack of explanation, ambiguous recommended reason, unrepresentative item description, low effectiveness, low efficiency, low user satisfaction, trust or willingness to accept the recommendation in most dialogue turns. 
Meanwhile, we label all sampled dialogues at the dialogue level and compute the correlation between dialogues and explanations.
This task covers effectiveness, efficiency, dialogue quality, and satisfaction~\cite{Jannach2020ASO}.
The results indicate that the explanation quality is highly correlated with the overall performance of the CRS.
Specifically, improving any perspectives of explanation is beneficial to the effectiveness and efficiency of CRS.
Dialogues with more diverse movie descriptions will be considered to have higher dialogue quality, while more trustworthy explanations lead to higher satisfaction.

\begin{figure}[t]
 \centering
 \includegraphics[width=1\columnwidth]{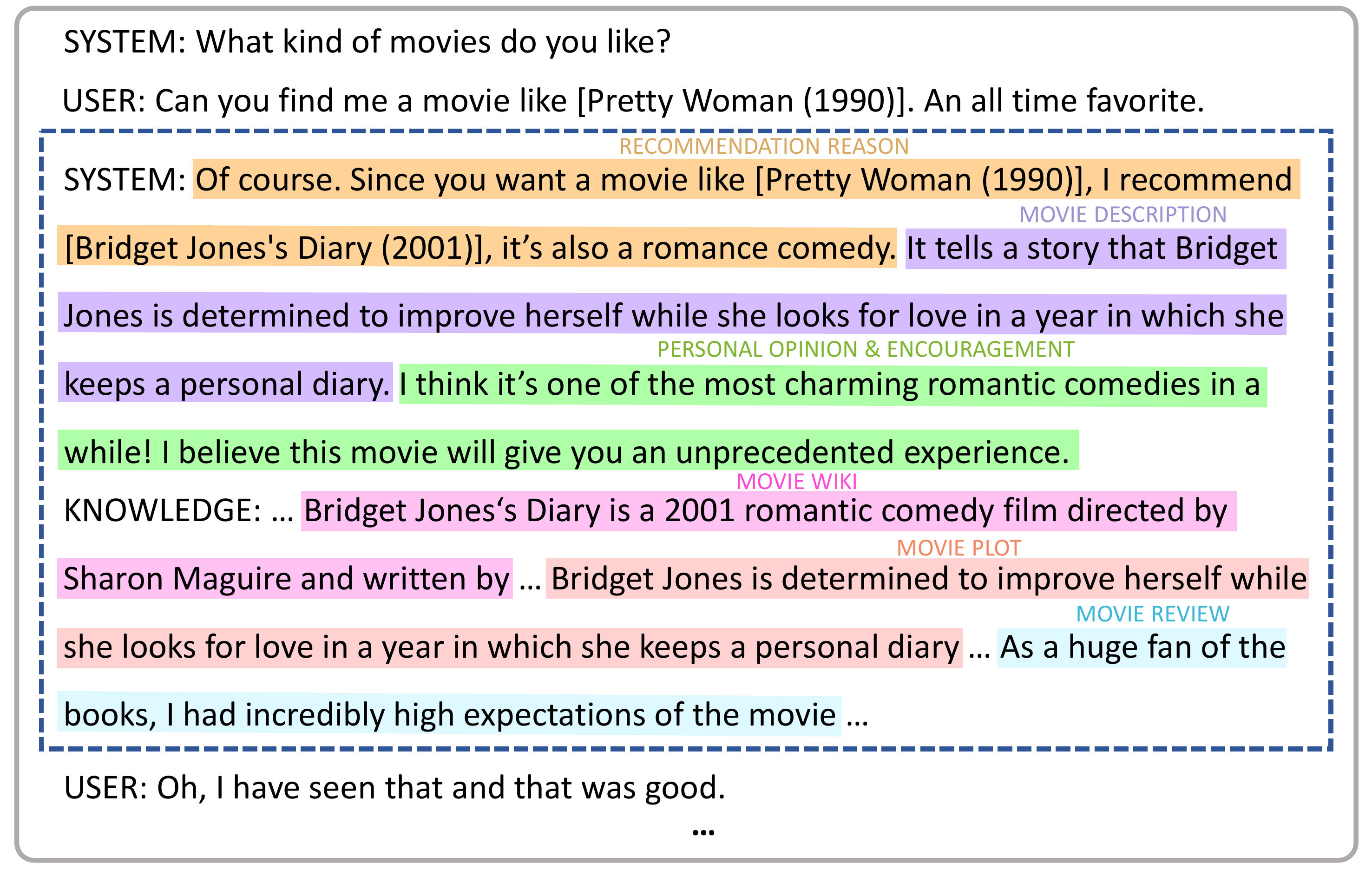}
 \caption{An example snippet of dialog in E-ReDial. Above each system's response with explanation and knowledge is different components. Best seen in colors.}
 \vspace*{-1.5mm}
 \label{figure:snip}
\end{figure}

To improve the explainability of CRS, we construct a new dataset named \textbf{E}xplainable \textbf{Re}commendation \textbf{Dial}ogues (\textbf{E-ReDial}).
We conduct a user study to investigate high-quality explanations for CRS and extract four main characteristics, i.e., clear recommendation reason, representative item description, encouragement or personal opinion, and reasonable and contextual.
Based on these, we conduct manual and automatic approaches to rewrite the system responses with low-quality explanations sampled from a commonly used real-world CRS dataset.
In the manual method, over 30 participants are involved. We ask them to search for relevant information about movies mentioned in responses (or knowledge for short), e.g., the plots, the trailers, the wikis and the reviews and extend responses to meet all requirements.
In the automatic method, we use \emph{GPT-3} for rewriting. We input the context with a prompt covering all requirements and collect the generated responses. 
We observe that the automatic method often fails to meet the rewriting requirements by evaluating the explanations obtained by both methods. Thus E-ReDial is constructed entirely by manual method.
It consists of 756 dialogues in the movie domain with 2,058 high-quality rewriting explanations, where each explanation is additionally annotated for corresponding knowledge and different components. Figure \ref{figure:snip} presents an illustrative example for our E-ReDial dataset.

We compare training-based and prompt-based approaches to perform explanation generation, i.e., given the context and the ground-truth recommended movies to generate corresponding explanations.
Experimental results suggest that models trained on E-ReDial can significantly improve explainability, while introducing knowledge into the models can further improve the performance.
In addition, both training-based and prompt-based methods have their own advantages. 
While GPT-3 can generate more realistic and diverse movie descriptions in an in-context learning setting. T5 training on E-ReDial can better generate clear reasons for recommendations based on user preferences.

In summary, this paper makes the following contributions: 
\begin{enumerate*}
    \item We propose the definition and evaluation perspectives of the explanation for CRS and verify its necessity.
    \item We collect and share a CRS dataset, E-ReDial, which includes over 2,000 high-quality explanations.
    \item We introduce both training-based and prompt-based baseline methods for explanation generation using E-ReDial.
\end{enumerate*}
The data and code are available at \url{https://github.com/Superbooming/E-ReDial}.



\section{Related work}
\label{sec:related_work}

\header{Conversational recommendation}
Unlike conventional recommendations, conversational recommenders interact with users through real-time, multi-turn conversations~\cite{Gao2021AdvancesAC}. Two main types of conversational recommender systems (CRS) are being studied: attribute-aware and topic-guided~\cite{Jannach2020ASO}. The former focus on the recommendation strategy~\citep{Zou2020NeuralIC, Luo2020DeepCF}, while the latter interacts with users through free-form natural language~\citep{Chen2019TowardsKR,Zou2020NeuralIC}.
We focus on topic-guided CRS in this paper. 
We analyze them from the view of explainability and enhance their corresponding abilities through manual and automatic methods.

\header{Explainable recommendation}
The explainable recommendation provides the items and explanations for why such items are recommended.
Providing explanations for the recommendations can help improve the recommender systems' transparency, persuasiveness, effectiveness, trustworthiness, and user satisfaction~\cite{Zhang2018ExplainableRA}. 
There are various forms of explanations, such as item features~\cite{chen2016learning, zhang2014explicit}, natural language~\cite{li2021extra, Geng2022ImprovingPE}, images~\cite{chen2019personalized},  and prototypes~\cite{Melchiorre2022ProtoMFPM}. 
Since CRS interacts with users through multi-turn conversations using natural language, we focus on the natural language explanations in this paper.
To evaluate the explainability of recommender system, \citet{Tintarev2015ExplainingRD} have proposed 7 evaluation goals: \emph{transparency}, \emph{scrutability}, \emph{trust}, \emph{effectiveness}, \emph{persuasiveness}, \emph{efficiency} and \emph{satisfaction}.
\citet{Chen2022MeasuringI} extend on these to propose a new goal, \emph{scrutability}, which has the same name mentioned above but has a different meaning. 
For the sake of distinction, we rename it as \emph{representativeness} in the remainder of this paper.
We integrate these goals into the CRS in Section \ref{subsec:evaluation}.

\begin{table*}[t]
    \small
    \centering
    \setlength\tabcolsep{9pt}
    \caption{Evaluation perspectives used in CRS. \textbf{D} denotes definition and \textbf{M} is for Measure method.}
    \vspace*{-1.5mm}
    \begin{tabular}{lll}
    \toprule
    \textbf{Metric} & \textbf{Definition and Measure Method} & \textbf{Reference} \\
    \midrule
    \emph{\textbf{Effectiveness}} & \textbf{D:} whether the explanation can help users to make good decisions~\cite{Tintarev2015ExplainingRD} & \cite{Tintarev2015ExplainingRD, Bilgic2005ExplainingRS, Chang2016CrowdBasedPN, Tintarev2008OverAU, Sharma2013DoSE} \\
    & \textbf{M:} difference between the ratings of the item before and after consuming & \\
    \midrule
    \emph{\textbf{Efficiency}} & \textbf{D:} whether the explanation can help users to make decisions faster~\cite{Tintarev2015ExplainingRD} & \cite{Tintarev2015ExplainingRD, Gedikli2014HowSI, Chang2016CrowdBasedPN} \\
    & \textbf{M:} time spent making decisions~\cite{Gedikli2014HowSI}, reading an explanation~\cite{Chang2016CrowdBasedPN}, the number of dialogue turns& \\
    \midrule
    \emph{\textbf{Persuasiveness}} & \textbf{D:} whether the explanation can convince users to accept the recommendation~\cite{Tintarev2015ExplainingRD}  & \cite{Tintarev2015ExplainingRD, Cosley2003IsSB} \\
    & \textbf{M:} accept rate of recommendation in conversations & \\
    \midrule
    \emph{\textbf{Transparency}} & \textbf{D:} whether the explanation explains how the system works~\cite{Tintarev2015ExplainingRD} & \cite{Tintarev2015ExplainingRD, Cramer2008TheEO} \\
    & \textbf{M:} user's perceived understanding of how the system works & \\
    \midrule
    \emph{\textbf{Scrutability}} & \textbf{D:} whether the explanation allows users to know their preference for the item~\cite{Tintarev2015ExplainingRD} & \cite{Tintarev2015ExplainingRD, Czarkowski2002ASA} \\
    & \textbf{M:} user's perceived understanding of how their preference is used to make recommendations & \\
    \midrule
    \emph{\textbf{Trust}} & \textbf{D:} whether the explanation increases users' confidence in the system~\cite{Tintarev2015ExplainingRD} & \cite{Tintarev2015ExplainingRD, Ohanian1990ConstructionAV} \\
    & \textbf{M:} related questions on transparency and scrutability & \\
    \midrule
    \emph{\textbf{Satisfaction}} & \textbf{D:} whether the explanation can make the use of the system enjoyable~\cite{Tintarev2015ExplainingRD} & \cite{Tintarev2015ExplainingRD, Symeonidis2008JustifiedRB, Felfernig2006AnES, McNee2003InterfacesFE,Sun2021SimulatingUS} \\
    & \textbf{M:} ask users whether they prefer a system with explanations~\cite{Symeonidis2008JustifiedRB} or whether they will choose the & \\
    & system again for the next & \\
    \midrule
    \emph{\textbf{Representativeness}} & \textbf{D:} whether the explanation can exactly correspond to the 
items~\cite{Chen2022MeasuringI} & \cite{Chen2022MeasuringI} \\
    & \textbf{M:} question-based & \\
    \midrule
    \emph{\textbf{Reasonability}} & \textbf{D:} whether the explanation is logically correct & - \\
    & \textbf{M:} question-based & \\
    \midrule
    \emph{\textbf{Coherence}} & \textbf{D:} whether the explanation is contextual & - \\
    & \textbf{M:} question-based & \\
    
    \bottomrule
    \end{tabular}
    \label{tab:eval_metric}
\end{table*}

\header{Explainable conversational recommendation}
There are only a few papers so far study explainable conversational recommendations. 
The existing works in CRS aim to make the generated responses more natural and fluent instead of more explanatory~\cite{DBLP:conf/acl/LuBSMCWH21, Zhou2020ImprovingCR, Ren2022VariationalRA}. 
Compared to single-turn explanation generation in conventional recommender systems, CRS generates multiple contextualized explanations, which poses further challenges for explanations. \citet{chen2021towards} is the first work on this topic. Still, it constructs dialogues using template-based user feedback to improve the performance of single-turn explanation generation rather than in a CRS scenario. 
\citet{Wen2022EGCREG} firstly propose a framework to generate explanations on a real-world CRS dataset. However, the performance of the generated explanations is limited and not effectively evaluated due to the lack of evaluation methods for CRS explanations and datasets with high-quality explanations. 
To alleviate these issues, we propose 10 evaluation perspectives and collect a new CRS dataset with high-quality explanations in this work.
\section{The explanation for CRS}
\label{sec:explainability}

In this section, we elucidate the meaning of the explanation for CRS, evaluation measurements and detail how to validate the necessity of explainability for CRS.

\subsection{Definition}
The definition of explanation in conventional recommender systems has been widely accepted as a justification for why items have been recommended~\cite{Zhang2018ExplainableRA}, or an item description helping the user better understand the qualities of the item at a broader level~\cite{Tintarev2015ExplainingRD}. 
CRS differs from conventional recommendation systems in many aspects. The major difference is that CRS dynamically interacts with users through real-time multi-turn conversations to elicit their preferences~\cite{Gao2021AdvancesAC}; thus, the scope of queries is enlarged. 
Combining these factors, we deem the explanation for CRS as:

\begin{description}
 \item[Explanation for CRS] \emph{A response from CRS that contains relevant details about the recommended items in the recommended round.}
\end{description}


\noindent
We expect the explanations only appear when CRSs are recommending items, i.e., responses in subsequent conversation rounds with details about the item can only be considered as an addition to its information and not as an explanation for the recommendation. 
In addition, an explanation must contain relevant details of recommended items, e.g., the recommendation reasons, the item descriptions, and personal opinions. 

\subsection{Evaluation perspectives}
\label{subsec:evaluation}

Explanations for recommendation can vary in purposes, e.g., providing users with recommendations and reasons to increase their trust in the system or providing detailed item descriptions to help users make a more informed decision. Thus, evaluating the explanations are expected to be comprehensive, i.e., covering different perspectives. We propose the following evaluation perspectives to better evaluate the explanation quality for CRS and list their definitions and evaluation methods in Table ~\ref{tab:eval_metric}.

Among these perspectives, the \emph{effectiveness}, \emph{efficiency}, \emph{persuasiveness}, \emph{satisfaction}, \emph{scrutability}, \emph{transparency}, \emph{trust}, and \emph{representativeness} are inherited from explainable recommender systems. 
They are used to evaluate properties common to all recommendation systems, e.g., system transparency, recommendation accuracy, and user satisfaction.
In addition, the \emph{reasonability} and \emph{coherence} is derived from the characteristics of the CRS.
They measure the consistency between explanations and conversations regarding logic and language, respectively.

\subsection{Necessity}
\label{subsec: necessity}

To validate the necessity of explainability for CRS, we look into five widely used CRS datasets as representatives and check their explanation quality at exchange and dialogue levels, respectively.


\header{Datasets}
The five CRS datasets are as follows. 
\begin{enumerate*}[label=(\arabic*)]
\item \emph{ReDial}~\cite{li2018conversational} is an English conversational movie recommendation dataset in realistic dialogue scenarios that contains 10,006 conversations.
\item \emph{TG-ReDial}~\cite{Zhou2020TowardsTC} is a Chinese conversational movie recommendation dataset incorporating topic threads annotated by semi-automatic approaches. It contains 10,000 conversations.
\item \emph{DuRecDial}~\cite{Liu2020TowardsCR} is a Chinese conversational recommendation dataset with multi-type dialogues which contains 10,190 conversations.
\item \emph{INSPIRED}~\cite{Hayati2020INSPIREDTS} is an English conversational movie recommendation dataset with good social strategies which contains 1,001 conversations.
\item \emph{OpenDialKG}~\cite{Moon2019OpenDialKGEC} is an English conversational recommendation dataset pairing each dialogue with corresponding knowledge graph paths, which contains 15,673 conversations.
\end{enumerate*}
%
We sample 50 dialogues randomly from each dataset for validation.

\begin{table*}[!t]
\centering
\setlength\tabcolsep{8pt}
\caption{
Annotation results resorting to evaluation perspectives we proposed in different CRS datasets.
The Explainable Rate ranges from 0 to 1, while the Overall metric ranges from 1 to 5. The other perspective metrics exhibit values between 1 to 3. Higher scores indicate better performance.
}
\vspace*{-1.5mm}
\label{table:datasets_explainability}
\begin{tabular}{l cc cc cc cc cc cc}
\toprule
\multirow{2}{*}{\textbf{Datasets}} 
& \multicolumn{2}{c}{\textbf{ReDial}} 
& \multicolumn{2}{c}{\textbf{TG-ReDial}}
& \multicolumn{2}{c}{\textbf{DuRecDial}}
& \multicolumn{2}{c}{\textbf{INSPIRED}}
& \multicolumn{2}{c}{\textbf{OpenDialKG}}
& \multicolumn{2}{c}{\textbf{E-ReDial (Ours)}}
\\

\cmidrule(lr){2-3} \cmidrule(lr){4-5}  \cmidrule(lr){6-7}  \cmidrule(lr){8-9} \cmidrule(lr){10-11} \cmidrule(lr){12-13}
& Avg     & Kappa
& Avg     & Kappa
& Avg     & Kappa
& Avg     & Kappa
& Avg     & Kappa
& Avg     & Kappa
\\ 

\midrule

Explainable Rate 
& 0.32 & 0.97 & 0.79 & 0.98 & 0.64 & 0.97 & 0.24 & 0.96 & 0.22 & - & \textbf{1.00} & 1.00 \\

Effectiveness
& 1.77 & 0.65 & 1.91 & 0.62 & 2.19 & 0.95 & 2.19 & 0.65 & 1.92 & - & \textbf{2.96} & 0.92\\

Efficiency
& 1.75 & 0.79 & 2.12 & 0.80 & 2.37 & 0.83 & 2.00 & 0.57 & 2.00 & - & \textbf{2.95} & 0.92\\

Persuasiveness
& 1.83 & 0.79 & 1.99 & 0.79 & 2.21 & 0.88 & 1.93 & 0.65 & 2.00 & - & \textbf{2.92} & 0.94\\

Transparency
& 1.79 & 0.78 & 1.97 & 0.70 & 2.21 & 1.00 & 1.89 & 0.54 & 1.92 & - & \textbf{2.97} & 0.97\\

Scrutability
& 1.69 & 0.79 & 1.94 & 0.65 & 2.07 & 0.91 & 1.78 & 0.60 & 2.00 & - & \textbf{2.95} & 0.93\\

Trust
& 1.90 & 0.71 & 2.08 & 0.82 & 2.26 & 0.85 & 2.07 & 0.64 & 2.00 & - & \textbf{2.89} & 0.93\\

Satisfaction
& 1.73 & 0.79 & 2.02 & 0.79 & 2.26 & 0.83 & 2.11 & 0.73 & 2.08 & - & \textbf{2.92} & 0.91\\

Representativeness
& 1.98 & 0.63 & 2.11 & 0.69 & 2.19 & 0.86 & 2.04 & 0.89 & 2.08 & - & \textbf{2.99} & 0.95\\

Reasonability
& 2.79 & 0.57 & 2.64 & 0.74 & 2.86 & 1.00 & 2.52 & 0.57 & 2.92 & - & \textbf{2.99} & 0.99\\

Coherence
& 2.54 & 0.68 & 2.55 & 0.83 & 2.84 & 0.92 & 2.48 & 0.65 & 2.75 & - & \textbf{2.99} & 0.99\\

\midrule

Overall
& 2.06 & 0.74 & 2.68 & 0.74 & 2.98 & 0.74 & 2.56 & 0.48 & 2.33 & - & \textbf{4.04} & 0.94\\

\bottomrule
\end{tabular}
\end{table*}

\header{Explanation quality at exchange level}
We design a questionnaire\footnote{See \url{https://github.com/Superbooming/E-Redial/blob/main/Exchange_level.md}} containing 13 questions covering all evaluation perspectives in Table~\ref{tab:eval_metric}.
Over 20 participants are invited to assess system responses from the sampled dialogues resorting to these evaluation metrics at the dialogue exchange level. 
For more details on annotation, see our repository \url{https://github.com/Superbooming/E-Redial/blob/main/Exchange_level.md}.
The annotation results are shown in Table \ref{table:datasets_explainability}. 
The metrics are obtained by averaging the corresponding perspectives' scores across all exchange turns.
We also compute Cohen's Kappa to measure the annotation disagreement.
Since there are few explanations in \emph{OpenDialKG}, its annotation results are highly agreed upon, so we omit Cohen's Kappa. 
Our results indicate that the existing datasets have relatively low-quality explanations for our proposed metrics. 
Specifically, lack of explanation, ambiguous recommended reason, unrepresentative item description, low effectiveness, low efficiency, low user satisfaction, trust, or willingness to accept the recommendation in most dialogue turns. 
The results differ across datasets, with the two Chinese datasets performing the best, \emph{INSPIRED} and \emph{OpenDialKG} the next best, and \emph{ReDial} the worst. 
The results also differ across metrics, \emph{reasonability} and \emph{coherence} are relatively high, while others are pretty low.

\begin{figure}[t]
 \centering
 \includegraphics[width=1\columnwidth]{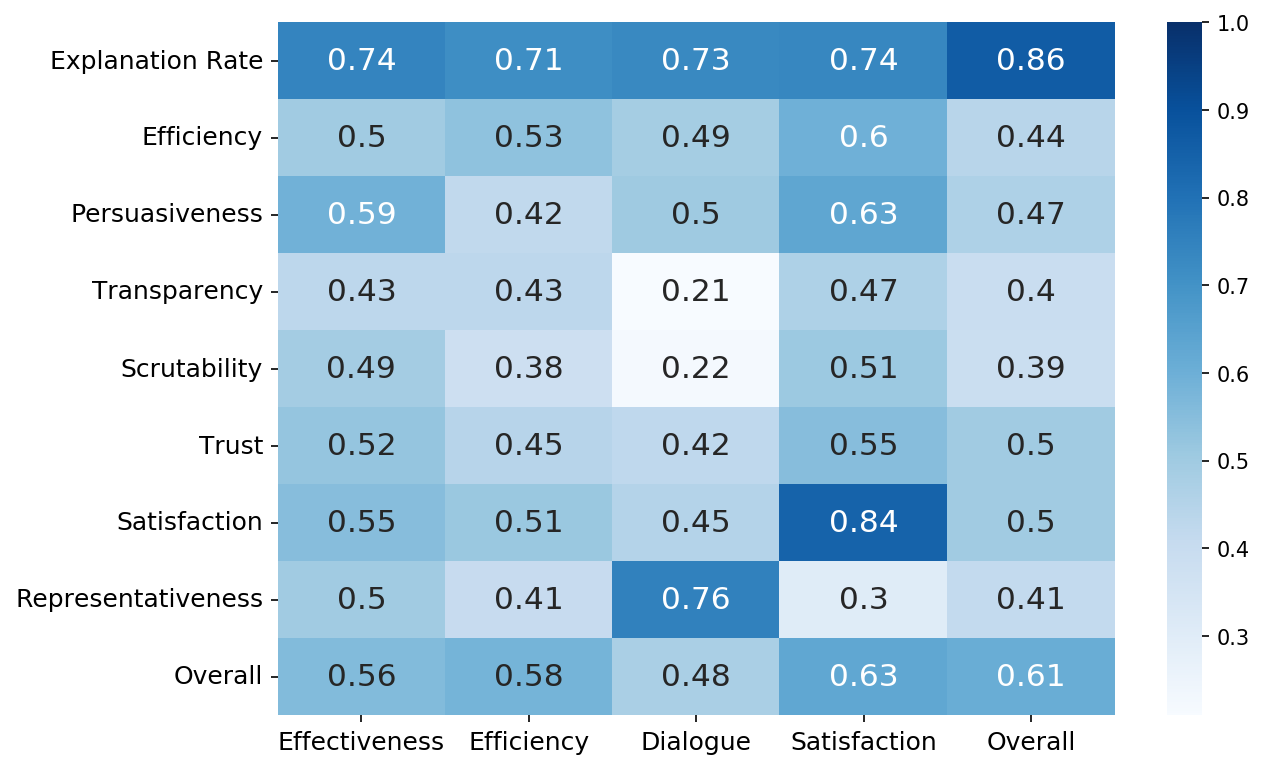}
 \caption{Spearman Correlation between the overall goals of CRS and different explanation perspectives. Each column corresponds to one CRS metric, and each row corresponds to one explanation metric.}
 \vspace*{-2mm}
 \label{figure:dia_exp}
\end{figure}

\header{Explanation quality at dialogue level}
We average the values on all explanation metrics in each dialogue as their overall ratings.
To figure out the relationship between explanation quality and dialogue performance, we ask annotators to label all sampled dialogues at the dialogue level.
We design another questionnaire\footnote{See \url{https://github.com/Superbooming/E-Redial/blob/main/Dialogue_level.md}} containing 5 questions covering the main overall goals of CRS, i.e., the \emph{effectiveness}, \emph{efficiency}, \emph{dialogue quality} and \emph{satisfaction}~\cite{Jannach2020ASO}.
These overall goals evaluate different aspects of the CRS, i.e., users' acceptance rate of recommendation, time cost to make decisions, the quality of conversation, and the system's success rate in meeting user needs. 
For more details on annotation, see our repository \url{https://github.com/Superbooming/E-Redial/blob/main/Dialogue_level.md}.
Based on the annotation result, we compute the Spearman Correlation between the explanation and CRS metrics. 
The correlation result is shown in Figure~\ref{figure:dia_exp}.
We omit the \emph{effectiveness}, \emph{reasonability} and \emph{coherence} of explanations since they are weakly correlated with CRS metrics.
Our results indicate that the explanation quality is highly correlated with the overall performance of the CRS. 
Specifically, the existence (\emph{explanation rate}) and quality of explanations (\emph{overall}) can significantly influence the performance of CRS.
The \emph{effectiveness} and \emph{efficiency} of CRS are strongly correlated with almost all explanation metrics, which means the improvement of any perspectives of explanation is beneficial to them.
The \emph{dialogue quality} of CRS is mainly related to \emph{representativeness} of explanation, which suggests that dialogues with more diverse movie descriptions will be considered to have higher dialogue quality. 
The \emph{satisfaction} of CRS has a higher correlation with trust-related metrics of explanation, e.g., the \emph{transparent} and \emph{scrutability}.
\section{Dataset Augmentation}
\label{sec:dataset}

This section elaborates on our efforts to improve the explainability of CRS. We conduct a user study to investigate the characteristics of good explanations for CRS and conduct both manual and automatic methods to rewrite the low-quality explanations on a commonly used CRS dataset.
Finally, we propose a new dataset with high-quality explanations, \emph{Explainable Recommendation Dialogues} (E-ReDial).
Below, we detail the creation of the dataset.

\subsection{User study} 
\label{subsec: user_study}


To explore the approaches to improving the explanation quality of CRS, we conduct a user study to investigate the characteristics of good explanations. Specifically, we invite over 20 participants and let each participant annotate 30 system responses containing explanations of different qualities, which we have evaluated in Section \ref{subsec: necessity}. 
We ask them to explain each response why the explanation is good or bad. 
By analyzing the explanations collected, we arrive at four main characteristics of high-quality explanations for CRS: 
\begin{enumerate*}[label=(\arabic*)]
    \item \emph{Clear recommendation reason} is a rational explanation about why the system recommends these items and what preferences the recommendation is based on.
    \item \emph{Representative item description}, which means the detailed information about recommended items, e.g., the directors, the actors, the genres, and the plot of recommended movies. The description should be true and have a cinematic characteristic.
    \item \emph{Encouragement or personal opinion on recommended items}, which means persuading users to accept recommendations from the view of language or personal experience.
    \item 
    \emph{Be reasonable and contextual}, meaning the explanations must be logically and linguistically consistent with the conversation.
\end{enumerate*}
We aspire to enhance the overall explanation quality of the dataset by generating better explanations, as the existing explanations within the dataset are commonly lacking in meeting the four aforementioned characteristics.

We consider three methods to generate good explanations: re-collecting dialogues and rewriting low-quality explanations manually and automatically.
Given the difficulty and cost of re-collecting, we choose the latter two and will explore the approach to re-collecting in future work.
This work extends the \emph{ReDial} and rewrites its low-quality explanations. 
We choose \emph{ReDial} since it is more commonly used and more realistic than a dataset constructed by some specific schemas.
Meanwhile, according to annotation results in Section \ref{subsec: necessity}, it performs the worst explanation quality.
We conduct manual and automatic approaches to rewrite its explanations and compare the two methods' performance.

\subsection{Manual method}
\label{subsec: manual}

\begin{figure*}[t]
 \centering
 \includegraphics[width=2\columnwidth]{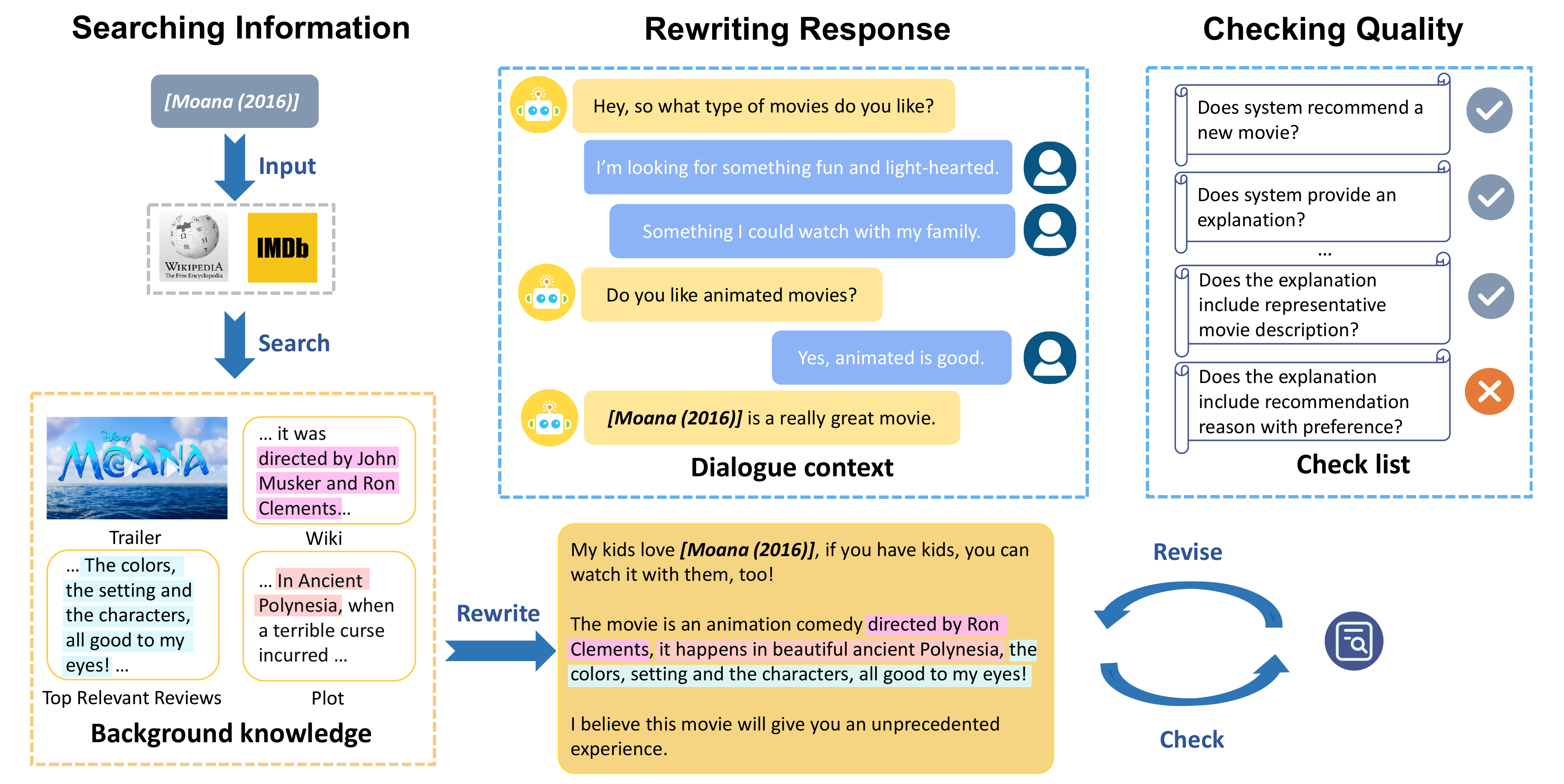}
 \caption{Illustration of the manual annotation pipeline. The response is structured into separate paragraphs for the recommendation reason, movie description, and encouragement, with different coloured highlights indicating varied information sources.}
 \label{figure:human}
\end{figure*}

Our manual method involves over 30 participants.
They are expected to search for relevant information and rewrite the response to meet all requirements in Section \ref{subsec: user_study}. 
The annotation pipeline includes three steps: searching for information, rewriting the response and checking quality.
Figure \ref{figure:human} illustrates the steps.

\noindent\textbf{Searching for information:}
To facilitate the needed information for rewriting the responses, such as the movie plot, the trailer, the reviews from \emph{IMDb}, and the wiki knowledge from \emph{Wikipedia}, our annotation pipeline provides the search function to retrieve them.
Specifically, the search function leverages \emph{Tf-IDF similarity} between the context and all reviews and returns the three most relevant reviews. 
For diversity, external information sources are also allowed.

\noindent\textbf{Rewriting the response:} 
The annotators are asked to rewrite the less explainable responses without changing the original meaning of the sentence. 
We further ask the annotators to label the rewritten responses' different aspects, i.e., the recommendation reasons, movie descriptions, encouragement or personal opinions.


\noindent\textbf{Checking quality:} 
Another 10 participants are instructed to check the qualify of the rewritten responses.
They are asked to answer the questions to determine whether the responses meet the requirements.
Responses having no qualified explanations are excluded and resent for annotations.

\subsection{Automatic method}

\begin{table}[!t]
\small
\centering
\setlength\tabcolsep{6pt}
\caption{
An example of extension methods. The automatic method misses [Meet the Parents (2000)] and fails to generate the description of [Meet the Fockers (2004)], while the manual method meets all requirements.
}
\label{table:human_auto}
\begin{tabular}{{@{}p{1\columnwidth}@{}}}
\toprule

\textbf{System:} Hi. \\

\textbf{User:} Hello! Could you recommend some comedy's? I've had a tough day. \\

\textbf{System:} Sure, have you seen [Meet the Parents (2000)]? It is soooo funny. \\

\textbf{User:} No, I haven't. \\

\midrule

\textbf{ReDial}: You have to see it. There is also [Meet the Fockers (2004)] which is the second part, they both are hilarious. \\

\midrule

\textbf{Automatic:} You should definitely watch [Meet the Fockers (2004)], since it's a hilarious comedy. It will surely make you laugh and bring some joy to your tough day. \\

\midrule

\textbf{Human:} Since you want some comedies, you should definitely watch these movies: [Meet the Parents (2000)], in this movie, male nurse Greg Focker meets his girlfriend's parents before proposing, but her suspicious father is every date's worst nightmare; [Meet the Fockers (2004)], which is the second part, this time, Focker came back again, and all hell breaks loose when the Byrnes family meets the Focker family for the first time! Both are hilarious, so don't miss them! They can surely bring you a happy day! \\

\bottomrule
\end{tabular}
\end{table}

In addition to the manual approach, we extend the explanations by applying automatic text generation models. 
We choose \emph{GPT-3}\footnote{\url{https://openai.com/api/}} as our benchmark model for an extension since it has achieved outstanding performance on lots of text generation tasks. 
We test two versions of \emph{GPT-3}, namely \code{text-davinci-003} and \code{code-davici-002}.
We input the dialogue context and the system response to be rewritten alongside a task instruction\footnote{The complete prompt we used is in \url{https://github.com/Superbooming/E-Redial/blob/main/Prompt.md}} covering all requirements in Section \ref{subsec: user_study}.
Then, we collect the generated text as the new response to the dialogue. 

We invite five annotators to check the quality of generated responses. 
Analyzing the results shows that more than 30\% of the generated responses need to meet the rewriting requirement. Specifically, changing the original meaning of the sentence, the lack of recommended movies, user preference, and movie descriptions in most cases. 
An example of extension results generated by the manual and automatic methods is shown in Table \ref{table:human_auto}. 

\subsection{The E-ReDial dataset}
After collecting all eligible manually rewritten responses from \emph{ReDial\footnote{The homepage of ReDial is \url{ https://redialdata.github.io/website/}. 
ReDial is published under the CC BY 4.0 License, allowing redistribution.}}, we put them together with the corresponding context to form new dialogues.
This results in a new CRS dataset, Explainable Recommendation Dialogues (E-ReDial).
It comprises 756 dialogues with 12,003 utterances, each with 15.9 turns on average. 
2,058 high-quality explanations are included, each with 79.2 tokens on average.
We annotate it in the same way as in Section \ref{subsec: necessity} to measure the explanation quality. 
The detailed evaluation results of E-ReDial are shown in Table \ref{table:datasets_explainability}.
The results show that our dataset is significantly better than others in all metrics.
\section{Experiments}
\label{sec:experiments}

\subsection{Dataset and metrics}
\header{Training and testing data}
We divide the newly labelled data into training and test sets at an 8:2 ratio. 
The test set comprises 150 conversations and 1121 system responses. 
Out of the system responses in the test set, 823 are idle or interrogative, with no movie recommendations, while the remaining 418 responses include movie recommendations and explanations.
We have categorized the subset of tests that contain recommendations and explanations as \emph{Test-Rec}, while the entire set of tests is referred to as \emph{Test-Full}.

\header{Automatic evaluation metrics}
Following previous studies on text generation, we utilize the following automatic evaluation metrics:
\emph{F1}, unigram F1 score that measures the similarity between the generated text and the ground-truth response.
\emph{BLEU}, we employ BLEU-2 (B2) and BLEU-4 (B4) implemented in the NLTK toolkit \url{https://www.nltk.org/}.
\emph{METEOR}, or MT, is a widely-used metric demonstrating good consistency in human evaluation.
\emph{Distinct} assesses text diversity by measuring the number of distinct n-grams in the generated text. 
Our experiments employ Distinct-2 (D2) and Distinct-3 (D3).

\header{Human Evaluation Metrics}
In addition to automatic evaluation, we conduct a human evaluation of various benchmark models. 
Specifically, we randomly sample 30 dialogues generated by the tested models and engage three well-informed annotators to evaluate the responses based on the metrics introduced in Section \ref{subsec:evaluation}.

\subsection{Benchmark models}
We test various benchmark dialogue generation models in E-ReDial:
\begin{itemize}[leftmargin=*,topsep=\parskip]
    \item \textbf{T5-Base~\cite{DBLP:journals/jmlr/RaffelSRLNMZLL20}}, a pre-trained encoder-decoder Transformer with 220M parameters;
    \item \textbf{BART-Base~\cite{Lewis2019BARTDS}}, a pre-trained encoder-decoder Transformer with 140M parameters;
    \item \textbf{GPT-2~\cite{radford2019language}}, a decoder-only Transformer with 117M parameters;
    \item \textbf{DialoGPT-Small~\cite{Zhang2019DIALOGPTL}}, a decoder-only language model that continues pre-trains GPT-2 on 2 billion dialogue corpus;
    \item \textbf{Flan-T5-XXL~\cite{Chung2022ScalingIL}}, an 11B-parameter language model that is finetuned on various NLP datasets using instruction tuning;
    \item \textbf{Davinci-002}, a large language model developed by OpenAI with about 170B parameters\footnote{We use the \code{code-davinci-002} API, and OpenAI has not disclosed the details of the parameters of the model until the time of writing the paper.} and the model is pre-trained on a mixed corpus including both code and text.
\end{itemize}
This paper aims to compare the efficacy of different approaches for generating responses with explanations rather than improving the accuracy of movie recommendations. 
Therefore, we keep the recommendation results as the ground truth for all benchmark models to ensure a fair comparison. 
The inputs for the models are the dialogue context and the names of the recommended movies, while the output is the response to the dialogue.

To further enhance the models, we propose a \textbf{knowledge-groun-} \textbf{ded (KG)} approach that utilizes the background knowledge of the recommended movies when generating recommendation explanations. 
For the KG approach, we incorporate each movie's plot, wiki, and reviews into the model input, as described in Section \ref{subsec: manual}. 
We only select the most relevant reviews based on the context to account for input length limitations.

Finetuning can be extremely costly in benchmark models incorporating large language models such as Flan-T5-XXL and Davinci-002. 
Therefore, we use a \textbf{prompt-based approach} with in-context learning (ICL). 
Specifically, we select eight (\emph{context}, \emph{response}) examples from the E-ReDial training set to serve as in-context exemplars, which are then inputted into the model alongside a task instruction\footnote{The complete prompt we used is in \url{https://github.com/Superbooming/E-Redial/blob/main/Prompt.md}}. 
Due to the model's limited context size, we do not input background knowledge to prompt-based models.

\header{Implementation details}
We implement the models using PyTorch and Huggingface Transformers.
All the finetuned models are optimized using AdamW optimizer with $lr{=}5e{-}5$, batch size of $16$, and are trained up to $20$ epochs.
During testing, for finetuned-based models, we employ greedy decoding.
For prompt-based models, i.e., Flan-T5-XXL and code-davinci-002, we employ nucleus decoding with top-$p{=}0.9$, tempearture${=}0.8$.

\subsection{Automatic evaluation results}

\begin{table}[!t]
\small
\centering
\setlength\tabcolsep{6pt}
\caption{
Performance for response generation among different training-based models in \emph{Test-Full}. Bold face indicates the best result in terms of the corresponding metric.
}
\vspace*{-1.5mm}
\label{table:results_model}
\begin{tabular}{l cccccc}
\toprule
& F1  & B2 & B4 & MT & D2 & D3 
\\ 



\midrule

T5-Base \\
+ ReDial 
& 20.35 & 3.65 & 1.70 & 11.34 & 34.04 & 39.23\\
+ ReDial + KG 
& 19.34 & 3.56 & 1.44 & 11.01 & 32.39 & 36.62\\
+ E-ReDial 
& 24.31 & 6.75 & 2.38 & 15.14 & 26.69 & 39.18\\
+ E-ReDial + KG
& \textbf{26.01} & \textbf{7.89} & \textbf{4.10} & \textbf{16.29} & \textbf{47.60} & \textbf{59.44}\\

\midrule
BART-Base\\
+ ReDial 
& 18.63 & 3.12 & 1.39 & 10.38 & 30.57 & 32.27\\
+ ReDial + KG 
& 18.04 & 2.71 & 1.37 & 9.86 & 31.91 & 32.97\\
+ E-ReDial 
& 23.63 & 6.26 & 2.52 & 14.11 & 26.96 & 35.60\\
+ E-ReDial + KG
& \textbf{24.46} & \textbf{7.87} & \textbf{4.61} & \textbf{16.48} & \textbf{47.38} & \textbf{59.11}\\

\midrule
GPT-2\\
+ ReDial 
& 19.82 & 3.61 & 1.54 & 11.80 & 28.10 & 34.51\\
+ ReDial + KG 
& 18.71 & 3.08 & 1.54 & 10.65 & 34.15 & 39.62\\
+ E-ReDial 
& 22.87 & 6.02 & 2.53 & 13.86 & 25.29 & 34.44\\
+ E-ReDial + KG
& \textbf{25.38} & \textbf{8.75} & \textbf{4.55} & \textbf{17.83} & \textbf{40.33} & \textbf{52.63}\\

\midrule
DialoGPT-Small\\
+ ReDial 
& 18.67 & 3.49 & 1.39 & 11.29 & 31.37 & 36.72\\
+ ReDial + KG 
& 18.12 & 3.11 & 1.46 & 10.33 & 29.34 & 33.77\\
+ E-ReDial 
& 22.80 & 6.06 & 2.13 & 14.23 & 19.02 & 26.83\\
+ E-ReDial + KG
& \textbf{25.70} & \textbf{8.12} & \textbf{4.22} & \textbf{16.61} & \textbf{30.87} & \textbf{40.34}\\

\bottomrule
\end{tabular}
\end{table}

\header{Performance of training-based models} 
Table \ref{table:results_model} displays the results of various training-based models for response generation. 
\emph{+ReDial} indicates that the models are trained on the original ReDial datasets, while \emph{+E-ReDial} indicates that they are trained on the newly collected E-ReDial dataset. 
The addition of \emph{+KG} indicates using a knowledge-grounded approach.

From the results, we have three key findings:
Firstly, the models trained on E-ReDial outperform those trained on ReDial across all backbone models and metrics.
This suggests that the proposed E-ReDial dataset can enhance the quality of CRS responses.
Secondly, incorporating knowledge-grounded methods (\emph{+KG}) significantly\footnote{Improvements are significant at $p<0.05$ level using t-test.} improves the performance of models trained on E-ReDial data, but no improvement is witnessed in the models trained on ReDial.
This may be because the ReDial responses merely mention the name of movies without providing explanations using background knowledge, unlike E-ReDial.
Lastly, the \emph{T5-Base + E-ReDial + KG} performs the best overall, and thus, we conduct a detailed analysis of it in our subsequent experiments.




\begin{table}[!t]
\small
\centering
\setlength\tabcolsep{5.75pt}
\caption{
Performance for response generation among models training on different data size in \emph{Test-Full}. Bold face indicates the best result in terms of the corresponding metric.
}
\label{table:results_data}
\begin{tabular}{l cccccc}
\toprule
& F1  & B2 & B4 & MT & D2 & D3 
\\ 

\midrule
\multicolumn{4}{l}{\emph{From T5-Base}}\\

+ E-ReDial (0\%)
& \phantom{0}9.74 & 1.70 & 0.40 & \phantom{0}6.17 & 27.80 & 36.12
\\

+ E-ReDial (5\%)
& 20.46 & 6.23 & 3.01 & 14.11 & 43.75 & 57.71
\\

+ E-ReDial (10\%)
& 21.46 & 6.92 & 3.69 & 15.46 & 41.29 & 52.52
\\

+ E-ReDial (50\%)
& 24.92 & \textbf{8.45} & \textbf{4.76} & \textbf{16.73} & 46.53 & 58.61
\\

+ E-ReDial (100\%)
& \textbf{26.01} & 7.89 & 4.10 & 16.29 & \textbf{47.60} & \textbf{59.44}
\\

\midrule
\multicolumn{4}{l}{\emph{From T5-Base finetuned on full ReDial}}\\

+ E-ReDial (0\%)
& 22.61 & 4.38 & 2.16 & 12.46 & 32.83 & 40.65
\\

+ E-ReDial (5\%)
& 25.50 & 7.92 & 3.90 & 17.12 & 42.92 & 57.68
\\

+ E-ReDial (10\%)
& 27.04 & 8.69 & 4.66 & 17.37 & 44.47 & 58.41
\\

+ E-ReDial (50\%)
& 27.63 & 9.65 & \textbf{5.00} & 18.65 & \textbf{47.72} & \textbf{60.28}
\\

+ E-ReDial (100\%)
& \textbf{27.98} & \textbf{9.71} & 4.84 & \textbf{18.88} & 47.07 & 59.20
\\

\bottomrule
\end{tabular}
\end{table}

\header{Impact of data size}
Since the proposed E-ReDial data size is relatively small, we conduct ablation experiments on data size to analyze its impact.
Table~\ref{table:results_data} presents the results of training the \emph{T5-Base+KG} model on different proportions of E-ReDial data, ranging from 0\% to 100\%.
We also compare the models finetuned from the vanilla T5 or the T5 pre-trained on ReDial data, with the former representing the first group in Table~\ref{table:results_data} and the latter representing the second group.
From the results, we see that as the amount of data increases, the models demonstrate improvement on all metrics until 50\% of the data is used. 
Beyond that point, the models show a decrease in some metrics. 
This finding suggests that the labelled data size of E-ReDial is sufficient to train the knowledge-grounded T5-Base model to generate explanations, and increasing the annotated data may provide limited gains.
Moreover, using the ReDial pre-trained model effectively enhances the model's performance in low-resource settings.
For example, using the pre-trained model, it is possible to use only 10\% of the E-ReDial data to outperform vanilla T5-Base trained with 100\% of the E-ReDial data.

\begin{table}[!t]
\small
\centering
\setlength\tabcolsep{3.5pt}
\caption{
Performance for response generation between training-based and prompt-based models in \emph{Test-Rec}. Bold face indicates the best result.
}
\label{table:results_prompt}
\begin{tabular}{l cccccc}
\toprule
& F1  & B2 & B4 & MT & D2 & D3 
\\ 

\midrule
T5-Base \emph{(+ReDial)}
& 14.76 & \phantom{0}0.47 & \phantom{0}0.21 & \phantom{0}8.37 & 41.70 & 48.74
\\

T5-Base \emph{(+E-ReDial)}
& 29.53 & 12.69 & \phantom{0}5.22 & 19.83 & 27.17 & 40.95\\


T5-Base \emph{(+E-ReDial +KG)}
& \textbf{34.62} & \textbf{18.10} & \textbf{11.22} & \textbf{26.65} & 48.66 & 62.24\\




\midrule

Flan-T5-XXL \emph{(8-shot ICL)} 
& 24.52 & \phantom{0}7.53 & \phantom{0}2.45 & 15.34 & 51.08 & 70.33\\
Davinci-002 \emph{(8-shot ICL)}
& 33.81 & 14.86 & \phantom{0}6.07 & 25.55 & \textbf{53.07} & \textbf{74.60}\\



\bottomrule
\end{tabular}
\end{table}

\header{Performance of prompt-based methods}
Table \ref{table:results_prompt} compares the results of two prompt-based models, Flan-T5-XXL and Davinci-002, with training-based T5 models.
We calculate the results only on \emph{Test-Rec} subset instead of the full test set. 
The results show that prompt-based models outperform training-based models in diversity (measured by D2 and D3).
Furthermore, Davinci-002 achieves comparable results to the best-performing training model T5-Base \emph{(+E-ReDial +KG)}, despite not explicitly inputting the background knowledge.

\begin{table}[!t]
\small
\centering
\setlength\tabcolsep{1.5pt}
\caption{
Human evaluation results of explanations generated by different models. Bold face indicates the best result in terms of the corresponding metric.
}
\label{table:results_human}
\begin{tabular}{l cccc}
\toprule
& T5+ReDial & T5+E-ReDial & T5+E-ReDial+KG & Davinci-002
\\ 

\midrule

Explanation rate & 0.23 & 1.00 & 1.00 & 1.00 \\ 

Effectiveness & 1.00 & 1.77 & 2.33 & \textbf{2.50} \\

Efficiency & 1.00 & 2.10 & 2.43 & \textbf{2.50}  \\

Persuasiveness & 1.00 & 2.00 & \textbf{2.30} & \textbf{2.30} \\

Transparency & 1.71 & \textbf{2.97} & \textbf{2.97} & 2.50 \\

Scrutability & 1.71 & 2.47 & \textbf{2.63} & 2.17 \\

Trust & 1.00 & 2.03 & \textbf{2.47} & 2.37 \\

Satisfaction & 1.14 & 2.00 & 2.47 & \textbf{2.57} \\

Representativeness & 1.43 & 2.10 & 2.70 & \textbf{2.83} \\

Reasonability & 1.71 & 2.60 & 2.73 & \textbf{2.77} \\

Coherence & 1.57 & 2.63 & 2.73 & \textbf{2.77} \\

Overall & 1.14 & 3.10 & \textbf{3.67} & 3.43 \\

\bottomrule
\end{tabular}
\end{table}

\begin{table*}[!t]
\small
\centering
\setlength\tabcolsep{6pt}
\caption{
Case study.
}
\vspace*{-2mm}
\label{table:case}
\begin{tabular}{{@{}p{1\textwidth}@{}}}
\toprule

\rowcolor{gray!30}[][] \textbf{Example 1}\\

\midrule

\textbf{User:} Do you recommend any movies? \\

\textbf{System:} Do you like Marvel movies? \\

\textbf{User:} Sure! What do you recommend to me?
\\

\midrule

\textbf{T5+ReDial:} [Spider-Man (2002)] is a good one.\\

\midrule

\textbf{T5+E-ReDial:} I will recommend [Spider-Man (2002)] since you like Marvel movies. It tells that Spider-Man, a mutant, must step forward to lead his people into a new future and must confront a challenger from his country's past. You will love this movie!
\\

\midrule

\textbf{T5+E-ReDial+KG:}
Since you like love movies, then I think these movies will suit your taste: [Spider-Man (2002)]. With great power comes great responsibility. These words, spoken by Peter Parker's uncle Ben (Cliff Robertson) during their last conversation and which have become Spider-Man's motto, also perfectly represent the challenge Sam Raimi was facing when asked to bring one of America's most loved superheroes to the screen. I believe that you will enjoy yourself while watching it.
\\

\midrule

\textbf{Davinci-002:}
I will recommend you [Spider-Man (2002)] since you like Marvel movies. After being bitten by a genetically altered spider, nerdy high school student Peter Parker is endowed with amazing powers. I will recommend you this movie and hope you like it.
\\

\midrule

\rowcolor{gray!30}[][] \textbf{Example 2}\\

\midrule

\textbf{User:} See any good films lately? Any dramas or comedies you'd lie to suggest?  
\\

\textbf{System:} I recommend [Wonder Woman (2017)] and [The Big Sick (2017)] because they are both drama or comedy films, [...]   
\\

\textbf{User:} I was dissapointed in the last two films I saw [The Shape of Water  (2017)] and [Lady Bird  (2017)]. haven't seen [Wonder Woman  (2017)] . Loved [The Big Sick (2017)]
\\

\midrule

\textbf{T5+ReDial:} I also recommend [Get Out (2017)], which is a drama film about a young girl who falls in love with a man who is a teen.  
\\

\midrule

\textbf{T5+E-ReDial:} [Get Out (2017)] is a good one. Since you want a good drama, [Get Out (2017)] is a good one. The story is that a young man who is trying to get out of a house with his girlfriend and her friend is forced to make a decision that will affect their future. I highly recommend it.
\\

\midrule

\textbf{T5+E-ReDial+KG:} Since you like dramas or comedies, then I think these movies well suit your taste [Get Out (2017)] It's nostalgic in its embracing of contemporary and classic horror and also brave in the accomplished handling of the film's combative themes. This movie is worth watching, you will like it. 
\\

\midrule
\textbf{Davinci-002:} [Get Out (2017)] is a great horror film. It's about Chris and his girlfriend meeting her parents for the first time. At first, Chris reads the family's overly accommodating behavior as nervous attempts to deal with their daughter's interracial relationship, but as the weekend progresses, a series of increasingly disturbing discoveries lead him to a truth that he could have never imagined. I recommend this movie to you, you won't be disappointed.
\\

\bottomrule
\end{tabular}
\end{table*}

\subsection{Human evaluation results}
Table \ref{table:results_human} shows the human evaluation results of explanations generated by four benchmark models. 
The results suggest that the T5 model trained on ReDial data (T5+ReDial) not only has a low explanation generation rate, accounting for only 23\% but produces low-quality responses, with an overall score of only 1.14.
Training the model on the proposed E-ReDial significantly improves the explanation generation rate and quality.
Furthermore, comparing the results of the explanations generated by T5-Base with and without knowledge-grounded generation indicates that introducing knowledge can substantially enhance almost all metrics.
This finding is consistent with the automatic evaluation results shown in Table \ref{table:results_model}
Additionally, T5-Base+KG and Davinci-002 exhibit their strengths, with T5-Base+KG having a slightly higher overall rating. 
By analyzing the generated responses, we infer that Davinci-002 generates more human-like and diverse movie descriptions, improving the effectiveness and efficiency of conversation and leading to higher user satisfaction and acceptance.
T5-Base+KG generates more precise reasons for recommendations based on user preferences, reasons increasing user trust and persuasiveness. 

\subsection{Case study}
Table~\ref{table:case} shows two examples of the model outputs.
In the first example, T5+ReDial recommends the movie without explanation. T5-E-ReDial explains, but its description of the movie is vague and unconvincing due to a lack of background knowledge.
T5-E-ReDial+KG, quoting classic scenes from the movie, better recommends the movie [Spider-Man (2002)] to users. 
Davinci-002 also explains the recommendations by acknowledging the user preferences and introducing the movie plot.
A similar phenomenon can be seen in the second example.
T5+ReDial gives a short explanation that is non-factual and unconvincing.
T5+E-ReDial's response is more explanatory but also includes factual mistakes.
T5+E-ReDial+KG and Davinci-002 generate responses that satisfy the users, whereas T5+E-ReDial+KG gives more apparent reasons based on user preferences, and Davinci-002's response is more realistic and contains diverse movie descriptions.

\section{Conclusion}
\label{sec:conclusion}
To elaborate on more explainable CRS, we have elucidated the meaning of the explanation for CRS and proposed ten evaluation perspectives to evaluate them. We further assessed existing benchmarks and verified the necessity of improving the explanation quality for CRS.
To improve the explanation quality, both manual and automatic methods are employed. Our manual method has collected and released a new CRS dataset, E-ReDial, which includes 756 dialogues with 2,058 high-quality explanations. In addition, we compared two baselines for explanation generation: training-based and prompt-based methods.
Experiments conducted on the newly collected dataset suggest that models trained on E-ReDial can significantly improve explainability while introducing knowledge into the models can further improve the performance.
GPT-3 in the in-context learning setting can generate more realistic and diverse movie descriptions. In contrast, T5 training on E-ReDial can better generate clear reasons for recommendations based on user preferences.

\header{Resource utilization}
The E-ReDial dataset can be used for explanation generation and other conversational information access tasks.
As a CRS dataset, it can be used for any CRS-related tasks.
Since each explanation is annotated with the knowledge, it can also be used in knowledge-grounded conversation.
The components annotated on explanations, i.e., the recommendation reasons, the items' descriptions, encouragement, or personal opinions, can be used for corresponding generation tasks.

\header{Limitations}
There are some limitations in this work.
Firstly, we improve the explainability of CRS by extending an existing CRS dataset.
We rewrite the responses with low-quality explanations via manual and automatic methods.
All rewritten responses passing the quality check will be collected, and we have put them with the corresponding context to construct our dataset.
Since the dialogue is obtained by extending the existing dialogue rather than from a real-world conversation scene, its explanations may not be realistic enough. 
Secondly, We only use the commonly used text generation models to generate explanations without designing a dedicated module, lacking further exploration.

\header{Future work}
This work opens up several directions for future studies in building explainable conversational recommender systems, as follows.
Firstly, we will explore more effective approaches to improving the explainability of CRS datasets, e.g., constructing dialogues from scratch. 
Secondly, we will investigate more complex evaluation methods for explanation quality, e.g., automatic metrics.
Finally, we will explore more methods to generate better explanations automatically.

\begin{acks}
This research was funded by the Natural Science Foundation of China (62272274,61972234,62072279,62102234,62202271), the Natural Science Foundation of Shandong Province (ZR2022QF004), the Key Scientific and Technological Innovation Program of Shandong Province (2019JZZY010129), Shandong University multidisciplinary research and innovation team of young scholars (No.2020QNQT017), the Fundamental Research Funds of Shandong University.
\end{acks}


\bibliographystyle{ACM-Reference-Format}
\balance
\bibliography{references}










\end{sloppypar}
\end{document}